\documentclass[12pt]{article}
\usepackage{hyperref}
\usepackage[english]{babel}
\usepackage{amsmath}
\usepackage{latexsym}
\usepackage{amssymb}
\usepackage[all]{xy}
\pdfoutput=1
\usepackage{graphicx}
\usepackage{psfrag}

\numberwithin{equation}{section} \numberwithin{table}{section}
\numberwithin{figure}{section}

\begin{document}

\begin{titlepage}

   \begin{center}

     \vspace{20mm}

     {\LARGE \bf Corner contributions to \\holographic entanglement entropy \vspace{5mm}\\in non-conformal backgrounds
     \vspace{3mm}}

     \vspace{10mm}

      Da-Wei Pang\\
     \vspace{5mm}
      {\small \sl Mathematical Sciences and STAG Research Center\\
      University of Southampton, Southampton SO17 1BJ, UK\\}
       {\small \tt d.pang\underline{~}at\underline{~}soton.ac.uk}
     \vspace{10mm}

   \end{center}

\begin{abstract}
\baselineskip=18pt
We study corner contributions to holographic entanglement entropy in non-conformal backgrounds: a kink for D2-branes as well as a cone and two different types of crease for D4-branes.
Unlike $2+1$-dimensional CFTs, the corner contribution to the holographic entanglement entropy of D2-branes exhibits a power law behaviour rather than a logarithmic term. However, the logarithmic term emerges in the holographic entanglement entropy of D4-branes. We identify the logarithmic term for a cone in D4-brane background as the universal contribution under appropriate limits and compare it with other physical quantities.
\end{abstract}
\setcounter{page}{0}
\end{titlepage}

\pagestyle{plain} \baselineskip=19pt

\tableofcontents

\section{Introduction}
The entanglement entropy (EE) has been playing an increasingly important role in a variety of research areas, including condensed matter physics~\cite{Amico:2007ag}, quantum information~\cite{Nielsen:2000qi}
and quantum field theory~\cite{Calabrese:2004eu, Casini:2009sr}. In quantum field theory the EE of a subsystem $A$ is defined by $S_{EE}=-{\rm Tr}(\rho_{A}\log\rho_{A})$, where $\rho_{A}$ is the reduced density matrix of $A$, given by tracing out the degrees of freedom in the complement of $A$ in the total density matrix. It is a formidable task to evaluate the EE in conventional framework, while an elegant method that relates the EE and area of minimal surface was proposed in~\cite{Ryu:2006bv, Ryu:2006ef}, dubbed as the holographic entanglement entropy (HEE). In the static case, the HEE of a subsystem $A$ is given by
\begin{equation}
\label{rtsec1}
S_{EE}=\frac{{\rm Area(\gamma)}}{4G_{N}},
\end{equation}
where $\gamma$ is the bulk minimal surface whose boundary is homologous to that of $A$. The formula~(\ref{rtsec1}) holds in the large $N$ limit, where the bulk theory is Einstein gravity coupled to various matter fields. Proofs of~(\ref{rtsec1}) were performed in~\cite{Casini:2011kv} for a spherical entangling region using conformal mapping and in~\cite{Lewkowycz:2013nqa} for a general entangling region using Euclidean quantum gravity. For reviews of HEE, see~\cite{Nishioka:2009un, Takayanagi:2012kg}.

The EE in the vaccum of a $d-$dimensional scale invariant field theory with a smooth entangling surface is given by~\cite{Liu:2012eea}
\begin{equation}
S_{EE}=\frac{H^{d-2}}{\delta^{d-2}}+\cdots+\frac{H}{\delta}+(-1)^{\frac{d-1}{2}}s_{d}+\frac{\delta}{H}+\cdots~~{\rm for~odd~d},
\end{equation}
\begin{equation}
S_{EE}=\frac{H^{d-2}}{\delta^{d-2}}+\cdots+\frac{H^{2}}{\delta^{2}}+(-1)^{\frac{d-2}{2}}s_{d}\log\frac{H}{\delta}+{\rm const}+\frac{\delta^{2}}{H^{2}}+\cdots~~{\rm for~even~d},
\end{equation}
where $H$ is the size of the entangling region and $\delta$ is the UV cutoff. The leading terms in both expressions exhibit the celebrated `area law'~\cite{Eisert:2008ur}, while $s_{d}$ is an $H$-independent constant, which gives the universal part of the EE. It is expected that the information of the subsystem is encoded in $s_{d}$.

If the entangling region is singular, new universal contributions to the EE will arise. For example, consider a three-dimensional CFT with the entangling region containing a single opening angle $\Omega$, the corresponding EE is given by~\cite{Hirata:2006jx}
\begin{equation}
S_{EE}=\frac{H}{\delta}-a\log(\frac{H}{\delta})+s_{3}+\mathcal{O}(\frac{\delta}{H}),
\end{equation}
where the coefficient of the new logarithmic term $a$ is a function of the opening angle $a=a(\Omega)$. Generalisations to higher-dimensional singular surfaces were extensively investigated in~\cite{Myers:2012vs}, where it was observed that more universal terms emerge as spacetime dimension increases. Recently the authors of~\cite{Bueno:2015rda, Bueno:2015xda} considered the universal corner contributions in $2+1$-dimensional CFTs in the presence of finite $N$ corrections as new measures of the degrees of freedom and compared those contributions with other physical quantities. In particular, they conjectured the following relation holds for any CFTs with holographic duals,
\begin{equation}
\frac{\kappa}{C_{T}}=\frac{\pi^{2}}{6}\Gamma\left(\frac{3}{4}\right)^{4},
\end{equation}
where $\kappa$ is determined by $a(\Omega)$ in the limit of $\Omega\rightarrow0$,
\begin{equation}
\lim_{\Omega\rightarrow0}a(\Omega)\approx\frac{\kappa}{\Omega},
\end{equation}
and $C_{T}$ is the central charge determined by the two-point function of the stress tensor for a $d$-dimensional CFT,
\begin{eqnarray}
& &\langle T_{ab}(x)T_{cd}(0)\rangle=\frac{C_{T}}{x^{2d}}\mathcal{I}_{ab,cd}(x),\nonumber\\
& &\mathcal{I}_{ab,cd}(x)\equiv\frac{1}{2}(I_{ac}(x)I_{db}(x)+I_{ad}(x)I_{cb}(x))-\frac{1}{d}\delta_{ab}\delta_{cd},\nonumber\\
& &I_{ab}(x)\equiv\delta_{ab}-2\frac{x_{a}x_{b}}{x^{2}}.
\end{eqnarray}
A more surprising conjecture for $2+1$-dimensional CFTs is
\begin{equation}
\label{sigmact}
\frac{\sigma}{C_{T}}=\frac{\pi^{2}}{24},
\end{equation}
where $\sigma$ is determined by $a(\Omega)$ in the $\Omega\rightarrow\pi$ limit,
\begin{equation}
\label{sigmasec1}
\lim_{\Omega\rightarrow\pi}a(\Omega)\approx\sigma(\pi-\Omega)^{2}.
\end{equation}
It has been shown that~(\ref{sigmact}) holds in both free field theories and holographic calculations~\cite{Bueno:2015rda, Bueno:2015xda, Elvang:2015jpa}. For recent developments see~\cite{Alishahiha:2015goa, Miao:2015dua, Bueno:2015qya, Bueno:2015lza}.

In this paper we study corner contributions to holographic entanglement entropy in non-conformal backgrounds, aiming at exploring if similar universal contributions emerge in non-conformal theories and to what extent such contributions provide measures of the degrees of freedom for the underlying quantum field theory. Our analysis is carried out in the lower-dimensional descriptions of non-conformal D2- and D4-branes, where these branes can be taken as exact solutions of Einstein-scalar theories. For D2-branes the singular entangling region is a kink $k$, as shown in the following figure, while for D4-branes three different singular entangling surfaces will be considered: a cone $c_{2}$, a crease $k\times\mathbb{R}^{2}$ and another crease $c_{1}\times\mathbb{R}^{1}$. Unlike $2+1$-dimensional CFTs, the universal logarithmic contribution to the EE of D2-branes does not exist, while for D4-branes logarithmic contributions of the form $\log(\delta/H)$ persist. The coefficients in front of such terms depend on both $\Omega$ and $H$, which is not surprising because conformal invariance is lost in this background. However, we still refer to the coefficients that are inversely proportional to $H$ as `universal', in a certain abuse of nomenclature. We also compare such universal contributions with other types of universal contributions arising from the EE of a strip and a sphere, as well as thermal entropy. We observe that comparing to the counterparts in $CFT_{5}$, the ratios of the universal contributions in D4-brane backgrounds take smaller values. This behaviour may be interpreted as the decrease of degrees of freedom along the RG flow.

The rest of this paper is organised as follows: in section 2 we briefly review some fundamental aspects of non-conformal D-branes, which also provide backgrounds for subsequent analysis. The corner contributions for D2-branes are evaluated in section 3, where we find the new term due to the corner exhibits a power law behaviour rather than logarithmic. The corner contributions for D4-branes are extensively discussed in section 4, including the cone $c_{2}$, the crease $c_{2}$, the crease $k\times\mathbb{R}^{2}$ and the crease $c_{1}\times\mathbb{R}^{1}$. In section 5 we identify universal contributions of the form $\log(\delta/H)$ and compare the corresponding coefficients with other physical quantities. Finally we summarise our results and discuss the implications in section 6.
\begin{figure}
\begin{center}
\includegraphics[angle=0,width=0.6\textwidth]{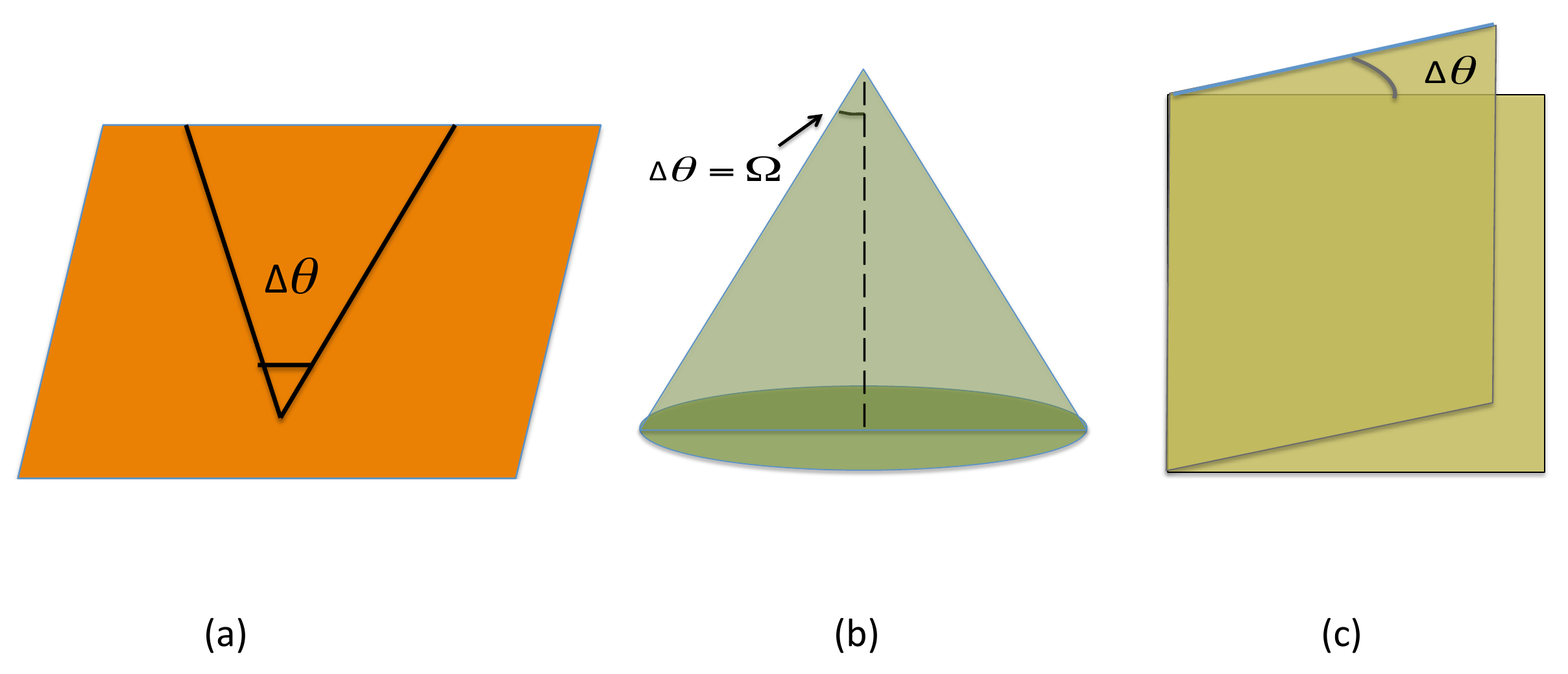}
\caption{Examples of singular surfaces: (a) a kink in $2+1$-dimensions; (b) a cone $c_{1}$; (c) a crease $k\times\mathbb{R}^{1}$.}
\label{fig:strip}
\end{center}
\end{figure}

\section{Non-conformal D-branes}
In this section we review some fundamental aspects of non-conformal D-branes for further discussions. The solutions describing $N$ coincident extremal Dp-branes in string frame are given by~\cite{Itzhaki:1998dd}
\begin{eqnarray}
ds^{2}_{\rm str}&=&H_{p}^{-1/2}(-dt^{2}+\sum\limits_{i=1}^{p}dx_{i}^{2})+H_{p}^{1/2}\sum\limits_{m=p+1}^{p}dx_{m}^{2},\nonumber\\
H_{p}&=&1+\frac{d_{p}g_{\rm YM}^{2}N}{\alpha^{\prime2}U^{7-p}},~~d_{p}=2^{7-2p}\pi^{\frac{9-3p}{2}}\Gamma(\frac{7-p}{2}),\nonumber\\
e^{\phi}&=&H_{p}^{(3-p)/4},
\end{eqnarray}
where $g_{\rm YM}$ denotes the Yang-Mills coupling. In the decoupling limit
\begin{equation}
\alpha^{\prime}\rightarrow0,~~U\equiv\frac{r}{\alpha^{\prime}}={\rm fixed},~~g_{\rm YM}^{2}={\rm fixed},
\end{equation}
the solutions become
\begin{eqnarray}
ds^{2}_{\rm str}&=&\alpha^{\prime}\left(\frac{U^{(7-p)/2}}{g_{\rm YM}\sqrt{d_{p}N}}(-dt^{2}+\sum\limits_{i=1}^{p}dx_{i}^{2})+\frac{g_{\rm YM}\sqrt{d_{p}N}}{U^{(7-p)/2}}dU^{2}+g_{\rm YM}\sqrt{d_{p}N}U^{(p-3)/2}d\Omega_{8-p}^{2}\right),\nonumber\\
e^{\phi}&=&(2\pi)^{2-p}g_{\rm YM}^{2}\left(\frac{g_{\rm YM}^{2}d_{p}N}{U^{7-p}}\right)^{(3-p)/4}.
\end{eqnarray}
Both the curvature and the dilaton should be small if the supergravity solutions are trustable, which leads to
\begin{equation}
\label{geff}
1\ll g_{\rm eff}^{2}\ll N^{4/(7-p)},~~g^{2}_{\rm eff}=g_{\rm YM}^{2}NU^{p-3}.
\end{equation}
For the cases of interest, the above conditions become
\begin{eqnarray}
& &g_{\rm YM}^{2}N^{1/5}\ll U\ll g_{YM}^{2}N~~{\rm for~D2-branes,}\nonumber\\
& &N^{-1}\ll g_{\rm YM}^{2}U\ll N^{1/3}~~{\rm for~D4-branes.}
\end{eqnarray}

Non-conformal Dp-branes can be effectively described as exact solutions of $p+2$-dimensional Einstein gravity coupled to scalar~\cite{Boonstra:1998mp, Kanitscheider:2008kd},
\begin{eqnarray}
\label{action}
S&=&\frac{N^{2}\Omega_{8-p}}{(2\pi)^{7}}\int d^{p+2}\sqrt{-g}\left[R-\frac{1}{2}(\partial\Phi)^{2}+V(\Phi)\right],\nonumber\\
V(\Phi)&=&\frac{1}{2}(9-p)(7-p)N^{-2\lambda/p}e^{a\Phi},\\
\Phi&=&\frac{2\sqrt{2(9-p)}}{\sqrt{p}(7-p)}\phi,~~~\lambda=\frac{2(p-3)}{7-p},~~a=-\frac{\sqrt{2}(p-3)}{\sqrt{p(9-p)}},\nonumber
\end{eqnarray}
where $\phi$ denotes the dilaton in ten dimensions. The above theory admits the following solutions
\begin{eqnarray}
ds^{2}_{p+2}&=&(Ne^{\phi})^{2\lambda/p}\left[\frac{u^{2}}{\mathcal{R}^{2}}(-\tilde{f}(u)dt^{2}+\sum_{i=1}^{p}dx_{i}^{2})+\frac{\mathcal{R}^{2}}{u^{2}\tilde{f}(u)}du^{2}\right],\nonumber\\
e^{\phi}&=&\frac{1}{N}(g_{\rm YM}^{2}N)^{\frac{7-p}{2(5-p)}}(\frac{u}{\mathcal{R}})^{\frac{(p-7)(p-3)}{2(p-5)}},\\
\tilde{f}(u)&=&1-(\frac{u_{0}}{u})^{\frac{2(7-p)}{5-p}},~~\mathcal{R}=\frac{2}{5-p}.\nonumber
\end{eqnarray}
For simplicity we will neglect all numerical factors from now on, and as a result, the corresponding D2-brane metric is given by
\begin{equation}
ds^{2}_{\rm D2}=\frac{1}{(g_{\rm YM}^{2}Nz)^{1/3}}\left[\frac{1}{z^{2}}(-dt^{2}+dx_{1}^{2}+dx_{2}^{2})+\frac{dz^{2}}{z^{2}}\right],
\end{equation}
while the D4-brane metric reads
\begin{equation}
\label{dgd4}
ds^{2}_{\rm D4}=\frac{\sqrt{g_{\rm YM}^{2}N}}{z^{1/2}}\left[\frac{1}{z^{2}}(-dt^{2}+\sum\limits_{i=1}^{4}dx_{i}^{2})+\frac{dz^{2}}{z^{2}}\right].
\end{equation}
Note that the solutions are conformal to $AdS_{p+2}$.
\section{Corner contributions for D2-branes}
In this section we consider corner contributions to the HEE in D2-brane background, whose dual field theory is $2+1$-dimensional supersymmetric Yang-Mills theory. We will see that unlike $2+1$-dimensional CFTs, the universal logarithmic term does not exist. Instead, the corner contribution exhibits a power-law behaviour, which disappears when the entangling region is smooth.

For convenience we rewrite the formula~(\ref{rtsec1}) as
\begin{equation}
S_{EE}=\frac{{\rm Area}(\gamma)}{4G_{N}}=4\pi KN^{2}{\rm Area}(\gamma),~~K=\frac{\Omega_{8-p}}{(2\pi)^{7}}.
\end{equation}
The corner in the boundary field theory is parameterised by $(\rho,\theta)$, where $\rho\in[0,H],~~\theta\in[-\Omega,\Omega]$. The corresponding minimal surface is also characterised by $(\rho,\theta)$ and the radial coordinate $z$ is parameterised by $z=z(\rho,\theta)$. Hence the induced metric reads
\begin{equation}
ds^{2}_{\rm ind}=\frac{1}{(g_{\rm YM}^{2}N)^{1/3}z^{7/3}}\left[(1+z^{\prime2})d\rho^{2}+(\rho^{2}+\dot{z}^{2})d\theta^{2}+2z^{\prime}\dot{z}d\rho d\theta\right].
\end{equation}
The HEE is given by
\begin{equation}
S_{EE}=\frac{4\pi KN^{5/3}}{g_{\rm YM}^{2/3}}\int d\rho d\theta\frac{1}{z^{7/3}}\sqrt{\rho^{2}+\rho^{2}z^{\prime2}+\dot{z}^{2}}.
\end{equation}
We can extract the equation that determines the minimal surface from the above expression,
\begin{eqnarray}
\label{eomd2}
& &\rho z(\rho^{2}+\dot{z}^{2})z^{\prime\prime}+\rho z(1+z^{\prime2})\ddot{z}+(\frac{7}{3}\rho+2zz^{\prime})\dot{z}^{2}\nonumber\\
& &+(\frac{7}{3}\rho+zz^{\prime})(1+z^{\prime2})\rho^{2}-2\rho z\dot{z}z^{\prime}\dot{z}^{\prime}=0,
\end{eqnarray}
where $z^{\prime}=\partial_{\rho}z,~~\dot{z}=\partial_{\theta}z$. Note that since the metric is conformal to $AdS$ and the boundary is flat, it is still appropriate to parameterise $z=\rho h(\theta)$ as in~\cite{Myers:2012vs}. As a result, the equation of motion~(\ref{eomd2}) becomes
\begin{equation}
\label{eomdd2}
h(1+h^{2})\ddot{h}+\frac{7}{3}\dot{h}^{2}+h^{4}+\frac{10}{3}h^{2}+\frac{7}{3}=0.
\end{equation}
To solve the equation of motion we impose the following boundary conditions: first we require $h\rightarrow0$ as $\theta\rightarrow\pm\Omega$; next we introduce $h_{0}$ such that $h(0)=h_{0}$ at $\theta=0$ and note that  $\dot{h}(0)=0$, which means that $h_{0}$ is the maximum value of $h(\theta)$; finally we introduce the UV cutoff at $z=\delta$ and cutoff $\epsilon$ in the angle such that at $z=\delta, h(\Omega-\epsilon)=\delta/\rho$.

It can be seen that the equation~(\ref{eomdd2}) admits a conserved quantity
\begin{equation}
K_{2}=\frac{(1+h^{2})^{7/6}}{h^{7/3}\sqrt{1+h^{2}+\dot{h}^{2}}}=\frac{(1+h_{0}^{2})^{2/3}}{h_{0}^{7/3}},
\end{equation}
which allows us to solve for $\dot{h}$ using the fact that $h$ decreases near the boundary and $\dot{h}$ should be negative
\begin{equation}
\label{hdotd2}
\dot{h}=-\frac{\sqrt{1+h^{2}}\sqrt{(1+h^{2})^{4/3}-h^{14/3}K_{2}^{2}}}{h^{7/3}K_{2}}.
\end{equation}
Therefore the HEE turns out to be
\begin{equation}
S_{EE}=\frac{8\pi KN^{5/3}}{g_{\rm YM}^{2/3}}\int^{H}_{\delta/h_{0}}\frac{d\rho}{\rho^{4/3}}\int^{\Omega}_{\epsilon}d\theta\frac{1}{h^{7/3}}\sqrt{1+h^{2}+\dot{h}^{2}}.
\end{equation}
In order to extract the divergent terms in the HEE, it would be convenient to rewrite the integral in terms of $h$
\begin{equation}
\label{seed2}
S_{EE}=\frac{8\pi KN^{5/3}}{g_{\rm YM}^{2/3}}\int^{H}_{\delta/h_{0}}\frac{d\rho}{\rho^{4/3}}\int^{\delta/\rho}_{h_{0}}dh\frac{1}{\dot{h}h^{7/3}}\sqrt{1+h^{2}+\dot{h}^{2}}.
\end{equation}
By using~(\ref{hdotd2}) one can expand the integrand of $h$ in the limit $h\rightarrow0$,
\begin{equation}
\frac{1}{\dot{h}h^{7/3}}\sqrt{1+h^{2}+\dot{h}^{2}}\simeq-\frac{1}{h^{7/3}}-\frac{1}{2}K_{2}h^{7/3}.
\end{equation}
Hence the double integral in~(\ref{seed2}) can be expressed as
\begin{equation}
\label{douintd2}
I_{1}-\int^{H}_{\delta/h_{0}}\frac{d\rho}{\rho^{4/3}}\int^{\delta/\rho}_{h_{0}}dh\frac{1}{h^{7/3}},
\end{equation}
where
\begin{equation}
I_{1}=\int^{H}_{\delta/h_{0}}\frac{d\rho}{\rho^{4/3}}\int^{\delta/\rho}_{h_{0}}dh\left(\frac{1}{h^{7/3}}+\frac{1}{\dot{h}h^{7/3}}\sqrt{1+h^{2}+\dot{h}^{2}}\right).
\end{equation}

Our next task is to figure out divergent terms in the HEE. It is easy to see that the divergent contributions in the second term of~(\ref{douintd2}) are given by
\begin{equation}
\frac{3H}{4\delta^{4/3}}-\frac{3}{h_{0}\delta^{1/3}}.
\end{equation}
To obtain the divergent terms in $I_{1}$ we first integrate $I_{1}$ by parts
\begin{equation}
I_{1}=-\frac{3}{H^{1/3}}\int^{\delta/H}_{h_{0}}J(h)dh-3\delta I_{2},
\end{equation}
where
\begin{eqnarray}
J(h)&=&\frac{1}{h^{7/3}}+\frac{1}{\dot{h}h^{7/3}}\sqrt{1+h^{2}+\dot{h}^{2}},\nonumber\\
I_{2}&=&\int^{H}_{\delta/h_{0}}\frac{d\rho}{\rho^{7/3}}J(h)\big|_{h=\delta/\rho}.
\end{eqnarray}
We can reexpress $I_{2}$ by introducing $q=\delta/\rho$, which leads to
\begin{equation}
I_{2}=-\frac{1}{\delta^{4/3}}\int^{\delta/H}_{h_{0}}dqq^{1/3}J(q).
\end{equation}
Then we expand the above integration around $\delta=0$,
\begin{equation}
I_{2}=-\frac{1}{\delta^{4/3}}\left[\int^{0}_{h_{0}}dqq^{1/3}J(q)+\frac{\delta}{H}q^{1/3}J(q)\big|_{q=\delta/H}+\cdots\right].
\end{equation}
Since $J(q)\sim q^{7/3}$, the divergent term in $I_{2}$ is given by
\begin{equation}
I_{2}=-\frac{1}{\delta^{4/3}}\int^{0}_{h_{0}}dqq^{1/3}J(q)-\frac{\delta^{7/3}}{H^{11/3}}.
\end{equation}
Finally the divergent terms in the HEE for D2-branes are collected as follows
\begin{equation}
\label{divd2}
S_{EE}\big|_{\rm div}=\frac{8\pi KN^{5/3}}{g_{\rm YM}^{2/3}}\left[\frac{3H}{4\delta^{4/3}}+\frac{3}{\delta^{1/3}}\left(\int^{0}_{h_{0}}dqq^{1/3}J(q)-\frac{1}{h_{0}}\right)\right].
\end{equation}
It can be seen that the first term in~(\ref{divd2}) also exists in smooth entangling regions~\cite{vanNiekerk:2011yi, Pang:2013lpa}, however the second term in~(\ref{divd2}) is a new divergent contribution, due to the existence of the corner. We may conclude that the universal logarithmic contribution in $2+1$-dimensional CFTs does not persist when conformal symmetry is broken. In the next section we will see that logarithmic contributions re-emerge in certain cases in D4-brane background.
\section{Corner contributions for D4-branes}
In the previous section we have seen that the universal corner contribution $\log(H/\delta)$ in $2+1$-dimensional CFTs does not exist in D2-brane background, in which conformal symmetry is broken. However, as we will see in this section, the same logarithmic contribution can be restored in D4-brane background. We will consider three different cases of corner contributions: a cone $c_{2}$, a crease $k\times\mathbb{R}^{2}$ and another crease $c_{1}\times\mathbb{R}^{1}$. The logarithmic contribution will appear in the first and the third case.
\subsection{Cone $c_{2}$}
In this case the induced metric is given by
\begin{equation}
ds^{2}_{\rm ind}=\frac{g_{\rm YM}\sqrt{N}}{z^{5/2}}\left[(1+z^{\prime2})d\rho^{2}+(\rho^{2}+\dot{z}^{2})d\theta^{2}+2\dot{z}z^{\prime}d\rho d\theta+\rho^{2}\sin^{2}\theta d\Omega_{2}^{2}\right],
\end{equation}
which results in the following form of the holographic entanglement entropy
\begin{equation}
\label{eed4c2}
S_{EE}=4\pi K\Omega_{2}g_{\rm YM}^{2}N^{3}\int d\rho d\theta \frac{\rho^{2}\sin^{2}\theta}{z^{5}}\sqrt{\rho^{2}+\rho^{2}z^{\prime2}+\dot{z}^{2}}.
\end{equation}
We can easily obtain the equation that determines the minimal surface from~(\ref{eed4c2}), which reads
\begin{eqnarray}
\label{eomd4c2}
& &\rho^{2}\sin\theta z(\rho^{2}+\dot{z}^{2})z^{\prime\prime}+\rho^{2}\sin\theta z(1+z^{\prime2})\ddot{z}-2\rho^{2}\sin\theta z\dot{z}z^{\prime}\dot{z}^{\prime}+5\rho^{2}\sin
\theta(\dot{z}^{2}+\rho^{2}(1+z^{\prime2}))\nonumber\\& &+2\cos\theta z\dot{z}(\dot{z}^{2}+\rho^{2}(1+z^{\prime2}))+\rho\sin\theta zz^{\prime}(4\dot{z}^{2}+3\rho^{2}(1+z^{\prime2}))=0.
\end{eqnarray}
Since the D4-brane metric is conformal to $AdS_{6}$, we can still parameterise $z(\rho,\theta)$ as $z=\rho h(\theta)$, hence the equation~(\ref{eomd4c2}) has a simpler form
\begin{eqnarray}
\label{eomd4c2h}
& &3h^{4}\sin\theta+5\sin\theta(1+\dot{h}^{2})+2h^{2}\sin\theta(4+\dot{h}^{2})\nonumber\\
& &+2\cos\theta h\dot{h}(1+h^{2}+\dot{h}^{2})+h(1+h^{2})\sin\theta\ddot{h}=0.
\end{eqnarray}
Under this parametrisation the HEE becomes
\begin{equation}
\label{heed4c2}
S_{EE}=8\pi K\Omega_{2}N^{3}g_{\rm YM}^{2}\int^{H}_{\delta/h_{0}}\frac{d\rho}{\rho^{2}}\int^{\delta/\rho}_{h_{0}}dh\frac{\sin^{2}\theta}{\dot{h}h^{5}}\sqrt{1+h^{2}+\dot{h}^{2}}.
\end{equation}

To extract the divergent parts of the HEE, we have to expand the integrand of $h$ near $h=0$. For later convenience let us introduce $y=\sin\theta$, which leads to
$$\dot{h}=\frac{\sqrt{1-y^{2}}}{y^{\prime}(h)},~~\ddot{h}=-\frac{1}{y^{\prime3}}(yy^{\prime2}+(1-y^{2})y^{\prime\prime}),$$
where the prime denotes partial derivative with respect to $h$. Then the equation~(\ref{eomd4c2h}) becomes
\begin{eqnarray}
& &h(1+h^{2})y(1-y^{2})y^{\prime\prime}-yy^{\prime}(5+2h^{2}+(5+8h^{2}+3h^{4})y^{\prime2})\nonumber\\
& &+hy^{2}(4+3(1+h^{2})y^{\prime2})-2h(1+(1+h^{2})y^{\prime2})+(5+2h^{2})y^{3}y^{\prime}-2hy^{4}=0,
\end{eqnarray}
and we can solve it perturbatively near $h=0$ with the following solution
\begin{equation}
\label{d4c2y}
y=\sin\Omega-\frac{1}{4}\cos\Omega\cot\Omega h^{2}-\frac{1}{128}(5+\cos2\Omega)\cot^{2}\Omega\csc\Omega h^{4}+\mathcal{O}(h^{6}).
\end{equation}
In order to obtain the behaviour of $\dot{h}$ near $h=0$, we introduce $f(h)=\dot{h}$, $\ddot{h}=ff^{\prime}$, hence the equation~(\ref{eomd4c2h}) takes the following form
\begin{eqnarray}
& &h(1+h^{2})yff^{\prime}+2h\sqrt{1-y^{2}}f^{3}+yf^{2}(5+2h^{2})\nonumber\\
& &+2h(1+h^{2})\sqrt{1-y^{2}}f+(5+8h^{2}+3h^{4})y=0,
\end{eqnarray}
and we can obtain the solution of $f$ near $h=0$
\begin{equation}
\label{d4c2f}
f=-\frac{2\tan\Omega}{h}+\frac{3}{4}\cot\Omega h+\cdots.
\end{equation}

Combining~(\ref{d4c2y}) and~(\ref{d4c2f}), the expansion of the integrand in~(\ref{heed4c2}) near $h=0$ turns out to be,
\begin{equation}
-\frac{\sin^{2}\Omega}{h^{5}}+\frac{3\cos^{2}\Omega}{8h^{3}}-\frac{\cos^{2}\Omega}{16h}+\mathcal{O}(h).
\end{equation}
We can separate the divergent parts of the HEE as follows
\begin{equation}
S_{EE}=8\pi K\Omega_{2}N^{3}g_{\rm YM}^{2}(I_{1}+I_{2}),
\end{equation}
where
\begin{eqnarray}
I_{1}&=&\int^{H}_{\delta/h_{0}}\frac{d\rho}{\rho^{2}}\int^{0}_{h_{0}}dhJ(h),\nonumber\\
J(h)&=&\frac{\sin^{2}\theta}{\dot{h}h^{5}}\sqrt{1+h^{2}+\dot{h}^{2}}+\frac{\sin^{2}\Omega}{h^{5}}-\frac{3\cos^{2}\Omega}{8h^{3}}+\frac{\cos^{2}\Omega}{16h},
\end{eqnarray}
and
\begin{equation}
I_{2}=\int^{H}_{\delta/h_{0}}\frac{d\rho}{\rho^{2}}\int^{\delta/\rho}_{h_{0}}dh\left(-\frac{\sin^{2}\Omega}{h^{5}}+\frac{3\cos^{2}\Omega}{8h^{3}}-\frac{\cos^{2}\Omega}{16h}\right).
\end{equation}
After some algebra we arrive at the divergent part from $I_{1}$
\begin{equation}
\frac{h_{0}}{\delta}\int^{0}_{h_{0}}dhJ(h),
\end{equation}
as well as the divergent terms from $I_{2}$,
\begin{equation}
\frac{H^{3}\sin^{2}\Omega}{12\delta^{4}}-\frac{\sin^{2}\Omega}{4h_{0}^{3}\delta}-\frac{3H\cos^{2}\Omega}{16\delta^{2}}+\frac{3\cos^{2}\Omega}{8h_{0}\delta}+\frac{\cos^{2}\Omega}{16H}\log\delta
+\frac{h_{0}\cos^{2}\Omega}{16\delta}.
\end{equation}
Finally we collect all the divergent terms
\begin{eqnarray}
\label{heedivd4c2}
S_{EE}\big|_{\rm div}&=&8\pi K\Omega_{2}N^{3}g_{\rm YM}^{2}\Big[\frac{H^{3}\sin^{2}\Omega}{12\delta^{4}}-\frac{3H\cos^{2}\Omega}{16\delta^{2}}+\frac{\cos^{2}\Omega}{16H}\log(\frac{\delta}{H})\nonumber\\
& &+\frac{1}{\delta}\left(\frac{h_{0}}{16}\cos^{2}\Omega+\frac{3\cos^{2}\Omega}{8h_{0}}-\frac{\sin^{2}\Omega}{3h_{0}^{3}}-h_{0}\int^{h_{0}}_{0}dhJ(h)\right)\Big],
\end{eqnarray}
where the logarithmic term $\log(\delta/H)$ is apparent. The first term in~(\ref{heedivd4c2}), which is proportional to $1/\delta^{4}$, also exists when the entangling region is smooth~\cite{vanNiekerk:2011yi, Pang:2013lpa}. The other terms, including the logarithmic one, are due to the presence of the corner. We will discuss the logarithmic term in more detail in the next section.
\subsection{Crease $k\times\mathbb{R}^{2}$}
In this case the induced metric is given by
\begin{equation}
ds^{2}_{\rm ind}=\frac{\sqrt{g_{\rm YM}^{2}N}}{z^{5/2}}\left[(1+z^{\prime2})d\rho^{2}+(\rho^{2}+\dot{z}^{2})d\theta^{2}+2z^{\prime}\dot{z}d\rho d\theta+dx_{3}^{2}+dx_{4}^{2}\right],
\end{equation}
which leads to the following expression of the HEE
\begin{equation}
\label{heed4k}
S_{EE}=4\pi KN^{3}{g_{\rm YM}^{2}}L^{2}\int d\rho d\theta\frac{1}{z^{5}}\sqrt{\rho^{2}+\rho^{2}z^{\prime2}+\dot{z}^{2}},
\end{equation}
where $x_{3,4}\in[0,L]$. We can derive the equation that determines the minimal surface from~(\ref{heed4k})
\begin{eqnarray}
\label{eomd4k}
& &\rho z(\rho^{2}+\dot{z}^{2})z^{\prime\prime}+\rho z(1+z^{\prime2})\ddot{z}+(5\rho+2zz^{\prime})\dot{z}^{2}\nonumber\\
& &+(5\rho+zz^{\prime})(1+z^{\prime2})\rho^{2}-2\rho z\dot{z}z^{\prime}\dot{z}^{\prime}=0.
\end{eqnarray}
Here we still parameterise $z=\rho h(\theta)$, which results in a simple form of~(\ref{eomd4k})
\begin{equation}
\label{eomd4kh}
h(1+h^{2})\ddot{h}+5\dot{h}^{2}+h^{4}+6h^{2}+5=0.
\end{equation}

It can be seen that the equation~(\ref{eomd4kh}) admits a conserved quantity
\begin{equation}
K_{4}=\frac{(1+h^{2})^{5/2}}{h^{5}\sqrt{1+h^{2}+\dot{h}^{2}}}=\frac{(1+h_{0}^{2})^{2}}{h_{0}^{5}}.
\end{equation}
Using the fact that $h$ decreases near the boundary and $\dot{h}$ should be negative, we obtain
\begin{equation}
\label{hdotd4}
\dot{h}=-\frac{\sqrt{1+h^{2}}\sqrt{(1+h^{2})^{4}-h^{10}K_{4}^{2}}}{h^{5}K_{4}},
\end{equation}
hence the HEE turns out to be
\begin{equation}
S_{EE}=8\pi KN^{3}{g_{\rm YM}^{2}}L^{2}\int^{H}_{\delta/h_{0}}\frac{d\rho}{\rho^{4}}\int^{\Omega}_{\epsilon}d\theta\frac{1}{h^{5}}\sqrt{1+h^{2}+\dot{h}^{2}}.
\end{equation}
To extract the divergent part of the HEE, we rewrite the integral over $h$
\begin{equation}
\label{heed4kh}
S_{EE}=8\pi KN^{3}{g_{\rm YM}^{2}}L^{2}\int^{H}_{\delta/h_{0}}\frac{d\rho}{\rho^{4/3}}\int^{\delta/\rho}_{h_{0}}dh\frac{1}{\dot{h}h^{5}}\sqrt{1+h^{2}+\dot{h}^{2}}.
\end{equation}
Using~(\ref{hdotd4}) the integrand of $h$ in~(\ref{heed4kh}) can be expanded as
\begin{equation}
\frac{1}{\dot{h}h^{5}}\sqrt{1+h^{2}+\dot{h}^{2}}\simeq-\frac{1}{h^{5}}-\frac{1}{2}K_{4}h^{5},
\end{equation}
hence the double integral in~(\ref{heed4kh}) becomes
\begin{equation}
\label{douintd4}
I_{1}-\int^{H}_{\delta/h_{0}}\frac{d\rho}{\rho^{4}}\int^{\delta/\rho}_{h_{0}}dh\frac{1}{h^{5}},
\end{equation}
where
\begin{equation}
I_{1}=\int^{H}_{\delta/h_{0}}\frac{d\rho}{\rho^{4}}\int^{\delta/\rho}_{h_{0}}dh\left(\frac{1}{h^{5}}+\frac{1}{\dot{h}h^{5}}\sqrt{1+h^{2}+\dot{h}^{2}}\right).
\end{equation}

We can easily obtain the divergent contributions in the second term of~(\ref{douintd4})
\begin{equation}
\frac{H}{4\delta^{4}}-\frac{1}{3h_{0}\delta^{3}}.
\end{equation}
To evaluate the divergent terms in $I_{1}$, we first integrate $I_{1}$ by parts, which gives
\begin{equation}
I_{1}=-\frac{1}{3H^{3}}\int^{\delta/H}_{h_{0}}J(h)dh-\frac{\delta}{3} I_{2},
\end{equation}
where
\begin{eqnarray}
J(h)&=&\frac{1}{h^{5}}+\frac{1}{\dot{h}h^{5}}\sqrt{1+h^{2}+\dot{h}^{2}},\nonumber\\
I_{2}&=&\int^{H}_{\delta/h_{0}}\frac{d\rho}{\rho^{4}}J(h)\big|_{h=\delta/\rho}.
\end{eqnarray}
Next let us introduce $q=\delta/\rho$ and rewrite $I_{2}$ as
\begin{equation}
I_{2}=-\frac{1}{\delta^{4}}\int^{\delta/H}_{h_{0}}dqq^{3}J(q).
\end{equation}
We can expand the above expression around $\delta=0$,
\begin{equation}
I_{2}=-\frac{1}{\delta^{4}}\left[\int^{0}_{h_{0}}dqq^{3}J(q)+\frac{\delta}{H}q^{3}J(q)\big|_{q=\delta/H}+\cdots\right].
\end{equation}
Since $J(q)\sim q^{5}$, we can easily arrive at the following result
\begin{equation}
I_{2}=-\frac{1}{\delta^{4}}\int^{0}_{h_{0}}dqq^{3}J(q)-\frac{\delta^{5}}{H^{9}}.
\end{equation}
Finally we collect all the divergent terms
\begin{equation}
\label{heed4kdiv}
S_{EE}\big|_{\rm div}=8\pi KN^{3}{g_{\rm YM}^{2}}L^{2}\left[\frac{H}{4\delta^{4}}+\frac{1}{3\delta^{3}}\left(\int^{0}_{h_{0}}dqq^{3}J(q)-\frac{1}{h_{0}}\right)\right].
\end{equation}
Comparing~(\ref{heed4kdiv}) with~(\ref{divd2}), we can see that the divergent part of the HEE for this case exhibits a similar structure as that for D2-branes: the first term in~(\ref{heed4kdiv}) also exists when entangling region is smooth, while the second term takes a power law behaviour, which is due to the presence of the corner.
\subsection{Crease $c_{1}\times\mathbb{R}^{1}$}
In this case the induced metric takes the following form
\begin{equation}
ds^{2}_{\rm ind}=\frac{g_{\rm YM}\sqrt{N}}{z^{5/2}}\left[(1+z^{\prime2})d\rho^{2}+(\rho^{2}+\dot{z}^{2})d\theta^{2}+2\dot{z}z^{\prime}d\rho d\theta+\rho^{2}\sin^{2}\theta d\phi^{2}+dx_{4}^{2}\right],
\end{equation}
in which the HEE becomes
\begin{equation}
\label{heed4c1}
S_{EE}=8\pi^{2}Kg_{\rm YM}^{2}N^{3}L\int d\rho d\theta \frac{\rho\sin\theta}{z^{5}}\sqrt{\rho^{2}+\rho^{2}z^{\prime2}+\dot{z}^{2}},
\end{equation}
where we have set $x_{4}\in[0,L]$. We can derive the following equation for the minimal surface from~(\ref{heed4c1})
\begin{eqnarray}
\label{eomd4c1}
& &\rho^{2}z(\rho^{2}+\dot{z}^{2})z^{\prime\prime}+\rho^{2}z(1+z^{\prime2})\ddot{z}-2\rho^{2}z\dot{z}z^{\prime}\dot{z}^{\prime}+5\rho^{2}(\dot{z}^{2}+\rho^{2}(1+z^{\prime2}))\nonumber\\
& &+z\left(\cot\theta\dot{z}^{3}+3\rho\dot{z}^{2}z^{\prime}+\rho^{2}\cot\theta\dot{z}(1+z^{\prime2})+2\rho^{3}z^{\prime2}(1+z^{\prime2})\right)=0.
\end{eqnarray}
We can still parameterise $z=\rho h(\theta)$ similar to previous examples, which leads to a simpler form for the equation
\begin{eqnarray}
\label{eomd4c1h}
& &h(1+h^{2})\ddot{h}+\cot{\theta}h\dot{h}^3+(5+h^{2})\dot{h}^{2}\nonumber\\
& &+\cot\theta h(1+h^{2})\dot{h}+2h^{4}+7h^{2}+5=0.
\end{eqnarray}
Furthermore the HEE can be expressed as
\begin{equation}
\label{heed4c1h}
S_{EE}=16\pi^{2}KN^{3}g_{\rm YM}^{2}L\int^{H}_{\delta/h_{0}}\frac{d\rho}{\rho^{3}}\int^{\delta/\rho}_{h_{0}}dh\frac{\sin\theta}{\dot{h}h^{5}}\sqrt{1+h^{2}+\dot{h}^{2}}.
\end{equation}

The subsequent analysis is parallel to that in section 4.1. First let us introduce $y=\sin\theta$, under which the equation~(\ref{eomd4c1h}) becomes
\begin{eqnarray}
& &h(1+h^{2})y(y^{2}-1)y^{\prime\prime}+yy^{\prime3}(5+7h^{2}+2h^{4})+h(y^{2}-1)^{2}\nonumber\\
& &+h(1+h^{2})(1-2y^{2})y^{\prime2}+(5+h^{2})y(1-y^{2})y^{\prime}=0.
\end{eqnarray}
We can solve the above equation perturbatively near $h=0$ and obtain the following solution
\begin{eqnarray}
\label{d4c1y}
y&=&\sin\Omega-\frac{1}{8}\cos\Omega\cot\Omega h^{2}+\frac{1}{1024}(5\cos2\Omega-19)\cot^{2}\Omega\csc\Omega h^{4}\nonumber\\
& &+\frac{3}{4096}\sin\Omega(-15+\csc^{2}\Omega+11\csc^{4}\Omega+3\csc^{6}\Omega)h^{6}\log h\nonumber\\
& &+y_{6}h^{6}+\mathcal{O}(h^{6}),
\end{eqnarray}
where $y_{6}$ is an undetermined constant. Next let us introduce $f(h)=\dot{h}$, $\ddot{h}=ff^{\prime}$, the equation~(\ref{eomd4c1h}) turns out to be
\begin{eqnarray}
& &h(1+h^{2})yff^{\prime}+h\sqrt{1-y^{2}}f^{3}+yf^{2}(5+2h^{2})\nonumber\\
& &+h(1+h^{2})\sqrt{1-y^{2}}f+(5+7h^{2}+2h^{4})y=0,
\end{eqnarray}
and the perturbative solution of $f$ near $h=0$ is given by
\begin{eqnarray}
\label{d4c1f}
f&=&-\frac{4\tan\Omega}{h}+\frac{7\csc\Omega+\sin\Omega}{8\cos\Omega}h+f_{3}h^{3}\nonumber\\
& &-\frac{9}{128}\tan\Omega(3+\csc^{2}\Omega)(5+3\csc^{2}\Omega)\log hh^{3}+\mathcal{O}(h^{3}),
\end{eqnarray}
where $f_{3}$ is again an undetermined constant.

Combining~(\ref{d4c1y}) and~(\ref{d4c1f}), the expansion of the integrand of $h$ in~(\ref{heed4c1h}) near $h=0$ turns out be
\begin{equation}
\frac{b_{1}}{h^{5}}+\frac{b_{3}}{h^{3}}+\frac{b_{1}}{h}+b_{h}h\log h+\mathcal{O}(h),
\end{equation}
where
\begin{eqnarray}
b_{5}&=&-\sin\Omega,~~b_{3}=\frac{3}{32}\cos\Omega\cot\Omega,\nonumber\\
b_{1}&=&\frac{3}{2048}\cos\Omega\cot\Omega(-19+3\csc^{2}\Omega),\\
b_{h}&=&\frac{3}{65536}(125-116\cos2\Omega+15\cos4\Omega)\csc^{2}\Omega\cot^{2}\Omega.\nonumber
\end{eqnarray}
Therefore the divergent parts of the HEE can be separated as
\begin{equation}
S_{EE}=16\pi^{2} KN^{3}g_{\rm YM}^{2}L(I_{1}+I_{2}),
\end{equation}
where
\begin{eqnarray}
I_{1}&=&\int^{H}_{\delta/h_{0}}\frac{d\rho}{\rho^{3}}\int^{0}_{h_{0}}dhJ(h),\nonumber\\
J(h)&=&\frac{\sin^{2}\theta}{\dot{h}h^{5}}\sqrt{1+h^{2}+\dot{h}^{2}}-\frac{b_{5}}{h^{5}}-\frac{b_{3}}{h^{3}}-\frac{b_{1}}{h},
\end{eqnarray}
and
\begin{equation}
I_{2}=\int^{H}_{\delta/h_{0}}\frac{d\rho}{\rho^{3}}\int^{\delta/\rho}_{h_{0}}dh\left(\frac{b_{5}}{h^{5}}+\frac{b_{3}}{h^{3}}+\frac{b_{1}}{h}\right).
\end{equation}
Note that since $$\int dh h\log h=-\frac{h^{2}}{4}+\frac{1}{2}h^{2}\log h\rightarrow0$$ as $h\rightarrow0$, it does not contribute to the divergence.

Next we can obtain the divergent part from $I_{1}$
\begin{equation}
\frac{h_{0}}{2\delta^{2}}\int^{0}_{h_{0}}dhJ(h),
\end{equation}
as well as the divergent terms from $I_{2}$
\begin{eqnarray}
& &-\frac{b_{5}H^{2}}{8\delta^{4}}+\frac{1}{2}b_{3}\frac{\log\delta}{\delta^{2}}-\frac{b_{1}}{2H^{2}}\log\delta\nonumber\\
& &+\frac{1}{\delta^{2}}\left(\frac{b_{5}}{4h_{0}^{2}}-\frac{1}{2}b_{3}\log(Hh_{0})+\frac{b_{3}}{4}-\frac{b_{1}}{4}h_{0}^{2}\right).
\end{eqnarray}
Finally we collect all the divergent terms
\begin{eqnarray}
\label{heed4c1div}
S_{EE}\big|_{\rm div}&=&16\pi^{2}KN^{3}g_{\rm YM}^{2}L\Big[-\frac{b_{5}H^{2}}{8\delta^{4}}+\frac{1}{2}b_{3}\frac{\log\delta}{\delta^{2}}-\frac{b_{1}}{2H^{2}}\log(\frac{\delta}{H})\nonumber\\
& &+\frac{1}{\delta^{2}}\left(\frac{b_{5}}{4h_{0}^{2}}-\frac{1}{2}b_{3}\log(Hh_{0})+\frac{b_{3}}{4}-\frac{b_{1}}{4}h_{0}^{2}+\frac{1}{2}h_{0}^{2}\int^{0}_{h_{0}}dhJ(h)\right)\Big].
\end{eqnarray}
Comparing~(\ref{heed4c1div}) with~(\ref{heedivd4c2}), we can find two similarities: the leading term is proportional to $1/\delta^{4}$, which also exists when the entangling region is smooth; the divergent part also contains a logarithmic term $\log(\delta/H)$. The divergent term proportional to $1/\delta^{2}$ in~(\ref{heedivd4c2}) remains in the present case with a more complicated coefficient, while the $1/\delta$ term in~(\ref{heedivd4c2}) disappears. More interestingly, the result~(\ref{heed4c1div}) contains a new divergent term $\log\delta/\delta^{2}$, which does not exist in~(\ref{heedivd4c2}).
\section{Universal contributions}
Our results in the previous section indicate for cone $c_{2}$ and crease $c_{1}\times\mathbb{R}^{1}$, the expected logarithmic term $\log(\delta/H)$ does exist. However, the coefficients in front of these terms are dependent on both $\Omega$ and the size of the system $H$. It is natural for the coefficients to possess dependence on $H$, as conformal symmetry is broken. However, here we still pick up the term proportional to $1/H$ as the universal contribution, in a certain abuse of nomenclature. The reason is that the universal contributions identified in previous results are also proportional to $1/H$~\cite{vanNiekerk:2011yi}. Moreover, the corresponding coefficient takes a finite value in the limit $\Omega\rightarrow0$. We also compare the universal contribution with that in the HEE of a strip, a sphere and the thermal entropy.
\subsection{Universal terms for D4-branes}
In~\cite{Bueno:2015rda, Bueno:2015xda} the authors considered $\kappa$ and $\sigma$ defined as follows as central charges
\begin{equation}
\lim_{\Omega\rightarrow0}a(\Omega)\approx\frac{\kappa}{\Omega},~~\lim_{\Omega\rightarrow\pi}a(\Omega)\approx\sigma(\pi/2-\Omega)^{2}.
\end{equation}
For non-conformal cases, even if we cannot find quantities that respect the above limits, we should pick up the one with better behaviour.
First let us recall the term proportional to $\log(\delta/H)$ when the entangling region is a cone $c_{2}$,
\begin{equation}
S_{EE~\rm univ}=q_{c_{2}}(\Omega)\log(\frac{\delta}{H}),~~q_{c_{2}}(\Omega)=\frac{1}{2H}\pi^{2}KN^{3}g_{\rm YM}^{2}\cos^{2}\Omega.
\end{equation}
If we take the same limits, we obtain
$$q_{c_{2}}(\Omega)\sim\frac{1}{H}~{\rm as}~\Omega\rightarrow0,~~q_{c_{2}}(\Omega)\sim(\frac{\pi}{2}-\Omega)^{2}~{\rm as}~\Omega\rightarrow\frac{\pi}{2}.$$
For the case of $c_{1}\times\mathbb{R}^{1}$, the logarithmic term is given by
\begin{equation}
S_{EE~\rm univ}=q_{c_{1}\times\mathbb{R}^{1}}(\Omega)\log(\frac{\delta}{H}),~~q_{c_{1}\times\mathbb{R}^{1}}(\Omega)=-8\pi^{2}KN^{3}g_{\rm YM}^{2}L\frac{b_{1}}{H^{2}},
\end{equation}
Upon taking those limits, the coefficient $b_{1}$ becomes
$$b_{1}\sim\frac{9}{2048\Omega^{3}}~{\rm as}~\Omega\rightarrow0,~~b_{1}\sim-\frac{3}{128}(\frac{\pi}{2}-\Omega)^{2}~{\rm as}~\Omega\rightarrow\frac{\pi}{2}.$$

From the above analysis we can see that although both of the coefficients of the logarithmic terms satisfy the desired property in the $\Omega\rightarrow\pi/2$ limit, however, neither of them satisfies the expected one in the $\Omega\rightarrow0$ limit. Furthermore, the coefficient for the case $c_{1}\times\mathbb{R}^{1}$ diverges as $\Omega\rightarrow0$, while that for $c_{2}$ remains constant. So we will take the term in $c_{2}$ as `universal'
\begin{equation}
\label{kappae}
S_{EE~\rm univ}=\kappa_{E}\log(\frac{\delta}{H}),~~\kappa_{E}=q_{c_{2}}(0)=\frac{\pi^{2}K}{2H}N^{3}g_{\rm YM}^{2},
\end{equation}
and will compare it with other quantities in the next subsection.

\subsection{Comparison with other quantities}
For convenience in this subsection we will express every quantity in terms of the dimensionless effective coupling~(\ref{geff}), so the universal contribution for $c_{2}$ in D4-brane backgrounds reads
\begin{equation}
\label{kpe}
\kappa_{E}=\frac{\pi^{2}}{2}KN^{2}g^{2}_{\rm eff}(H^{-1}).
\end{equation}
The first quantity that we would like to compare is the universal contribution to the HEE of a strip in D4-brane background, where the corresponding induced metric is given by
\begin{equation}
ds^{2}_{\rm ind}=\frac{\sqrt{g_{\rm YM}^{2}N}}{z^{5/2}}\left[(1+x^{\prime2})dz^{2}+\sum\limits_{i=2}^{4}dx_{i}^{2}\right].
\end{equation}
Here we have set $x_{1}\equiv x(z)\in[-l/2,l/2],~x_{i}\in[0,H], i=2,3,4$.
The HEE is expressed as
\begin{equation}
S_{EE}=8\pi KN^{3}g_{\rm YM}^{2}H^{3}\int dz\frac{1}{z^{5}}\sqrt{1+x^{\prime2}},
\end{equation}
which admits the following conserved quantity
\begin{equation}
\frac{x^{\prime}}{z^{5}\sqrt{1+x^{\prime2}}}=\frac{1}{z_{t}^{5}},
\end{equation}
where $z_{t}$ denotes the turning point of the minimal surface in the bulk spacetime. Introducing $t=z/z_{t}$, we can express the boundary separation length
\begin{equation}
\frac{l}{2}=\int^{1}_{0}dt\frac{t^{5}}{\sqrt{1-t^{10}}}
\end{equation}
and the HEE
\begin{equation}
S_{EE}=8\pi KN^{3}g_{\rm YM}^{2}\frac{H^{3}}{z_{t}^{4}}\int^{1}_{0}dt\frac{1}{t^{5}\sqrt{1-t^{10}}}
\end{equation}
in terms of Gamma functions. Then the universal contribution is given by~\cite{vanNiekerk:2011yi, Pang:2013lpa}
\begin{equation}
\label{kappastrip}
S_{EE~\rm univ}=-\kappa_{\rm strip}(\frac{H}{l})^{4},~~\kappa_{\rm strip}=32\pi^{7/2}kN^{2}g_{\rm eff}^{2}(H^{-1})\left(\frac{\Gamma(\frac{3}{5})}{\Gamma(\frac{1}{10})}\right)^{5}.
\end{equation}
Comparing~(\ref{kappastrip}) with~(\ref{kpe}) we arrive at
\begin{equation}
\frac{\kappa_{\rm strip}}{\kappa_{E}}=2^{6}\pi^{3/2}\left(\frac{\Gamma(\frac{3}{5})}{\Gamma(\frac{1}{10})}\right)^{5}\approx0.0334931.
\end{equation}

The second quantity is the universal contribution to the HEE of a sphere with the following induced metric
\begin{equation}
ds^{2}_{\rm ind}=\frac{\sqrt{g_{\rm YM}^{2}N}}{z^{5/2}}\left[(1+\dot{\rho}^{2})dz^{2}+\rho^{2}d\Omega_{3}^{2}\right],
\end{equation}
where $\rho\in[0,H]$ and $\rho(0)=H$. The HEE is given by
\begin{equation}
S_{EE}=4\pi K \Omega_{3}N^{3}g_{\rm YM}^{2}\int dz\frac{\rho^{3}}{z^{5}}\sqrt{1+\rho^{\prime2}},
\end{equation}
which leads to the following equation of motion
\begin{equation}
-3z(1+\rho^{\prime2})+\rho(-5\rho^{\prime}-5\rho^{\prime3}+z\rho^{\prime\prime})=0.
\end{equation}
In order to extract the divergent terms in the HEE, we need the solution near $z=0$, which is
\begin{equation}
\rho(z)\approx H-\frac{3z^{2}}{8H}-\frac{45z^{4}}{512H^{3}}-\frac{45}{4096H^{5}}z^{6}\log z+\rho_{6}z^{6}+\cdots.
\end{equation}
This result enables us to expand the integrand in the HEE
\begin{equation}
\frac{H^{3}}{z^{5}}-\frac{27H}{32z^{3}}+\frac{135}{2048 Hz}+\cdots,
\end{equation}
and obtain the universal contribution~\cite{vanNiekerk:2011yi, Pang:2013lpa}
\begin{equation}
-\kappa_{\rm sphere}\log\frac{\delta}{H},~~\kappa_{\rm sphere}=\frac{45}{128}\pi^{2}KN^{2}g_{\rm eff}^{2}(H^{-1}).
\end{equation}
So the ratio of $\kappa_{\rm sphere}$ over $\kappa_{E}$ reads
\begin{equation}
\frac{\kappa_{\rm sphere}}{\kappa_{E}}=\frac{45}{64}=0.703125.
\end{equation}

The third quantity we would compare is the thermal entropy, which can be evaluated in near extremal black D4-brane background
\begin{equation}
ds^{2}=\frac{\sqrt{g_{YM}^{2}N}}{z^{5/2}}\left[-f(z)dt^{2}+\frac{dz^{2}}{f(z)}+\sum\limits_{i=1}^{4}dx_{i}^{2}\right],~~f(z)=1-\frac{z^{6}}{z_{0}^{6}}.
\end{equation}
We can easily obtain the temperature and entropy
\begin{equation}
T=\frac{3}{2\pi z_{0}},~~s=4\pi KN^{3}g_{\rm YM}^{2}z_{0}^{-5},
\end{equation}
which leads to the following relation
\begin{equation}
s=c_{E}T^{4},~~c_{E}=4\pi K N^{2}g_{\rm eff}^{2}(T)(2\pi/3)^{5}.
\end{equation}
Therefore we obtain the ratio of $c_{E}$ over $\kappa_{E}$
\begin{equation}
\frac{c_{E}}{\kappa_{E}}=\frac{2^{8}\pi^{4}}{3^{5}}\frac{g_{\rm eff}^{2}(T)}{g_{\rm eff}^{2}(H^{-1})}\approx102.62\frac{g_{\rm eff}^{2}(T)}{g_{\rm eff}^{2}(H^{-1})}.
\end{equation}
Note that since the black D4-brane is near extremal, the temperature is very low such that $g_{\rm eff}^{2}(T)/g_{\rm eff}^{2}(H^{-1})$ takes a very small value, while the constraint~(\ref{geff}) still holds.

For comparison, let us the counterparts for $CFT_{5}$. When the entangling region is $c_{2}$, the universal contribution to the HEE is given by~\cite{Myers:2012vs},
\begin{equation}
S^{\rm log}_{5}\Big|_{c_{2}}=16KN^{2}\pi^{2}q_{5}(\Omega)\log\left(\frac{\delta}{H}\right),
\end{equation}
where $q_{5}(\Omega)$ is determined by numerical fit,
\begin{equation}
\log|q_{5}|\approx-\log\sin\Omega-2.1.
\end{equation}
In the limit $\Omega\rightarrow0$, we can extract the universal contribution
\begin{equation}
S_{5, \rm univ}=\frac{\kappa_{5}}{\Omega}\log\left(\frac{\delta}{H}\right),~~\kappa_{5}\approx1.96KN^{2}\pi^{2}.
\end{equation}
On the other hand, the universal contributions to the EE of $CFT_{5}$ for a strip and a sphere are~\cite{Ryu:2006ef}
\begin{equation}
S_{5, \rm strip}=-\kappa_{5,\rm strip}\left(\frac{H}{l}\right)^{3},~~\kappa_{5,\rm strip}=\frac{2^{6}}{3}KN^{2}\pi^{3}\left(\frac{\Gamma(\frac{5}{8})}{\Gamma(\frac{1}{8})}\right)^{4},
\end{equation}
\begin{equation}
S_{5, \rm sphere}=\kappa_{5,\rm sphere}=4KN^{2}\pi^{5/2}\Gamma(-\frac{3}{2}),
\end{equation}
while the entropy density and temperature obeys
\begin{equation}
s_{5}=c_{5,E}T^{4},~~c_{5,E}=\frac{(4\pi)^{5}}{5^{4}}KN^{2}.
\end{equation}
Hence we arrive at the following ratios for $CFT_{5}$,
\begin{equation}
\frac{\kappa_{5,\rm strip}}{\kappa_{5}}\approx0.0449457,~~\frac{\kappa_{5, \rm sphere}}{\kappa_{5}}\approx8.54855,~~\frac{c_{5,E}}{\kappa_{5}}\approx25.9187.
\end{equation}
It can be seen that the first two ratios are larger than the counterparts for D4-branes. For the third ratio, note that it is reasonable to take $g_{\rm eff}^{2}(T)/g_{\rm eff}^{2}(H^{-1})\ll 1$, so the third ratio is also larger than that for D4-branes.

A possible interpretation for this behaviour may be as follows: we consider the action~(\ref{action}) as the IR limit of AdS gravity coupled to relevant deformation $\Phi$. The asymptotic UV geometry is still AdS while the IR geometry is given by~(\ref{dgd4}). As claimed in~\cite{Bueno:2015xda}, these universal quantities can characterise the degrees of freedom of the system, so it may be expected that the ratios also do the job. Our results indicate that as we go from the UV to the IR, the values of the ratios decrease, which might be seen as an analog of the celebrated c-theorem.  
\section{Summary and discussion}
New universal contribution of the form $\log(\delta/H)$ to the HEE arising from singular surfaces has been observed in various dimensional CFTs, as investigated in~\cite{Myers:2012vs, Bueno:2015rda, Bueno:2015xda}. One may wonder if terms of the same form still emerge when the background is non-conformal, and if they do exist, to what extent they are `universal'. In this paper we study corner contributions to the HEE in non-conformal D-brane backgrounds. To be specific, we focus on a kink $k$ in D2-brane background, as well as a cone $c_{2}$, a crease $k\times\mathbb{R}^{2}$ and another crease $c_{1}\times\mathbb{R}^{1}$ in D4-brane background. We observe that unlike $2+1$-dimensional CFTs, the logarithmic contribution disappears for D2-branes, whose dual field theory is $2+1$-dimensional supersymmeric Yang-Mills theory. There exists a new divergent term due to the presence of the corner, which exhibits a power law behaviour instead of logarithm.

Furthermore, the logarithmic term emerges in D4-brane background for the case of a cone $c_{2}$ and a crease $c_{1}\times\mathbb{R}^{1}$. However, the coefficients in front of such terms are dependent on both the angle $\Omega$ and the size of the subsystem $H$. We explore the behaviour of the coefficients in the limit of $\Omega\rightarrow0$ and $\Omega\rightarrow\pi/2$ respectively and find that even though both of them obey the expected property, being proportional to $(\pi/2-\Omega)^{2}$ as $\Omega\rightarrow\pi/2$, the coefficient for $c_{1}\times\mathbb{R}^{1}$ diverges as $1/\Omega^{3}$ and that for $c_{2}$ becomes a constant in the limit of $\Omega\rightarrow0$. Neither of them shares the property for CFTs that it is proportional to $1/\Omega$ as $\Omega\rightarrow0$. We take the logarithmic contribution for $c_{2}$ as `universal' in the sense that it is proportional to $1/H$, which agrees with other universal contributions to the HEE identified in previous investigations. We also compute the ratios of the coefficient of the universal term with that of the HEE of a strip, a sphere and the thermal entropy in D4-brane background, and compare the results with those for $CFT_{5}$. The values are smaller that those for CFT, which might be viewed as an analog of the c-theorem.

One may relate the existence of logarithmic terms in D4-brane background to its connection with M5-branes, since D4-branes are obtained by dimensional reduction in one of the worldvolume directions of M5-branes. Therefore it would be reasonable to expect that the HEE of D4-branes inherits some features of that of M5-branes. Actually in~\cite{Myers:2012vs} it was observed that for a crease $c_{2}\times\mathbb{R}^{1}$, the logarithmic term takes the following form
\begin{equation}
S^{\rm log}_{c_{2}\times\mathbb{R}^{1}}\propto\frac{(7-9\cos2\Omega)\cot^{2}\Omega}{256H}\log(\frac{\delta}{H}),
\end{equation}
which possesses a similar form as~(\ref{kappae}). If $\mathbb{R}^{1}$ is taken as the compactification direction of M5-branes, it may be possible that the HEE shares the same structure in the resulting D4-brane background. On the other hand, if we dimensionally reduce M2-branes along one of the worldvolume directions, we arrive at fundamental strings rather than D2-branes, hence the HEE would exhibit a different behaviour.

Our discussions are purely holographic in the sense that all the quantities were calculated in the dual gravity background. It would be interesting to see if the universal contribution can be re-derived from field theory considerations. Very recently exact results for corner contributions to the EE and Renyi entropies of free bosons and fermions in $2+1$ dimensions were obtained in~\cite{Elvang:2015jpa}, where it was found that the quantity $\sigma$ defined in~(\ref{sigmasec1}) and the central charge $C_{T}$ obeys
\begin{equation}
\frac{\sigma}{C_{T}}=\frac{\pi^{2}}{24},
\end{equation}
which agrees with the conjecture proposed in~\cite{Bueno:2015rda, Bueno:2015xda}. Therefore it would be desirable to investigate higher dimensional backgrounds and check if a similar relation can be derived, in particular in non-conformal backgrounds. A plausible way is to generalise the analysis in~\cite{Safdi:2012sn} and study subsequent implications on D4-branes.

\bigskip \goodbreak \centerline{\bf Acknowledgments}
\noindent
We would like to thank Ben Withers for helpful discussions, as well as Rob Myers and William Witczak-Krempa for correspondence.
DWP is supported by a Marie-Curie Intra-European Fellowship.

\newpage

\end{document}